\documentclass{ws-procs9x6}
\usepackage{epsf}
\usepackage{epsfig}
\usepackage{soul}
\usepackage{amsmath}
\usepackage{amssymb}
\usepackage{amsbsy}
\usepackage{amsfonts}
\usepackage{graphics}
\usepackage{latexsym}
\usepackage{cite}
\usepackage[mathscr]{eucal}
\newcommand{\bc}{\begin{center}}
\newcommand{\ec}{\end{center}}
\newcommand{\be}{\begin{equation}}
\newcommand{\ee}{\end{equation}}
\newcommand{\bea}{\begin{eqnarray}}
\newcommand{\eea}{\end{eqnarray}}

\newcommand{\eps}{\epsilon}

\def\Nf{n_f}

\newcommand{\Dlr}{\overset{\leftrightarrow}{D}}

\begin{document}

\title{
Soffer Bound and Transverse Spin Densities from Lattice QCD
\footnote{\uppercase{P}reprint \uppercase{E}dinburgh 2005/19, \uppercase{DESY} 05-218,
\uppercase{LTH} 680}
}

\author{M. Diehl$^{1}$, M. G\"ockeler$^2$, Ph.~H\"agler$^3$, R.~Horsley$^4$, D.~Pleiter$^5$,
P.E.L. Rakow$^6$, A. Sch\"afer$^2$,
G.~Schierholz$^{1,5}$ and J.M. Zanotti$^4$}

\address{
$^1$Deutsches Elektron-Synchrotron DESY,
22603 Hamburg, Germany\\
$^2$Institut f\"ur Theoretische Physik\\
Universit\"at Regensburg, 93040 Regensburg, Germany\\
$^3$Institut f\"ur Theoretische Physik T39,
Physik-Department der TU M\"unchen,
James-Franck-Stra\ss{}e,
D-85747 Garching, Germany\\
%
$^4$School of Physics,
University of Edinburgh,
Edinburgh EH9 3JZ, UK\\
$^5$John von Neumann-Institut f\"ur Computing NIC / DESY\\
15738 Zeuthen, Germany\\
$^6$Theoretical Physics Division,
Dep.~of Math.~Sciences,
University of Liverpool,
Liverpool L69 3BX, UK}

\author{QCDSF/UKQCD collaborations}

\maketitle

\abstracts{Generalized transversity distributions encode essential
information on the internal structure of hadrons related to transversely
polarized quarks. Lattice QCD allows us to compute the lowest
moments of these tensor generalized parton distributions. 
In this talk, we discuss a first lattice study of the Soffer bound and show 
preliminary results for transverse spin densities
of quarks in the nucleon.}

\section{Introduction}
Generalized parton distributions (GPDs) \cite{GPD,Diehl:2003ny} are an ideal tool
to study many fundamental facets of hadron structure
in terms of quarks and gluons. One key point is the relation of
GPDs to (orbital) angular momentum, which plays a central role for the
nucleon spin sum rule \cite{Ji:1996ek}.
Moreover, GPDs allow us to investigate the nontrivial
interplay of longitudinal momentum and transverse coordinate space
degrees of freedom \cite{Burkardt:2000za,Diehl:2002he,MIT-2}.
In this contribution, we will focus our attention on the
recently observed correlation of transverse quark spin and
impact parameter which shows up in transverse spin densities of quarks in the
nucleon \cite{Diehl:2005jf}. It turns out that these correlations in the transverse plane
are governed by quark helicity flip (or tensor) GPDs \cite{Diehl:2001pm}.
As in the case of the unpolarized and the polarized GPDs \cite{Diehl:2003ny}, they are defined
via the parametrization of an off-forward nucleon matrix element of a bilocal quark operator 
as follows
\begin{eqnarray}
\!\!\!&&\!\!\!\!\! \left\langle
   P',\Lambda ' \right|\int_{-\infty}^{\infty} \frac{d \lambda}{4 \pi} e^{i \lambda x}
  \bar q\!\left(-\frac{\lambda}{2}n\right)
  \, n_\mu\sigma^{\mu\nu}\gamma_5
  \mathcal{U}q\!\left(\frac{\lambda}{2} n\right)
  \left| P,\Lambda \right\rangle = \overline u(P',\Lambda ') n_\mu\bigg\{  \nonumber\\
   & & \sigma^{\mu\nu}\gamma_5
   \bigg( H_T(x, \xi, t) - \frac{t}{2m^2}\widetilde H_T(x, \xi, t) \bigg)
  + \frac{\eps^{\mu\nu\alpha\beta} \Delta_{\alpha} \gamma_{\beta}} {2 m} \overline E_T(x, \xi, t)  \nonumber\\
 &+&  \frac{\Delta^{[\mu} \sigma^{\nu]\alpha}\gamma_5 \Delta_{\alpha}} {2m^2 } \widetilde H_T(x, \xi, t)
 + \frac{\eps^{\mu\nu\alpha\beta} \overline P_{\alpha} \gamma_{\beta}} {m} \widetilde E_T(x, \xi, t)\!\bigg\}   
 u(P,\Lambda)
 \label{GPDs1}\, ,
\end{eqnarray}
where $f^{[\mu\nu]}=f^{\mu\nu}-f^{\nu\mu}$, $\Delta=P'-P$ is the momentum transfer with
$t=\Delta^2$, $\overline P = (P'+P)/2$, and $\xi=-n\cdot \Delta/2$
defines the longitudinal momentum transfer with the light-like vector $n$.
The Wilson line ensuring gauge invariance of the bilocal operator is denoted by $\mathcal{U}$.
%
Our parametrization in Eq.~(\ref{GPDs1}) in terms of the four independent tensor GPDs slightly differs
from the literature \cite{Diehl:2001pm,Diehl:2005jf} where a function $E_T$ instead of $\overline E_T$ has been used.
However, in \cite{Diehl:2005jf,Burkardt:2005hp} it has been noted that $E_T$ typically appears in
linear combination with the tensor GPD $\widetilde H_T$. It is therefore 
reasonable to adopt a new notation and consider $\overline E_T =  E_T+2\widetilde H_T$
and $\widetilde H_T$ as fundamental quantities.

One prominent feature of GPDs is that they reproduce the well known
parton distributions in the forward limit, $\Delta=0$, and that their
integral over the momentum fraction $x$ leads directly to form factors.
For this reason, the GPD $H_T(x,\xi,t)$ is called
generalized transversity, since for vanishing momentum transfer
it is equal to the transversity parton distribution, $H_T(x,0,0)=\delta q(x)=h_1(x)$ for
$x>0$ and $H_T(x,0,0) = -\delta\bar{q}(-x)=-\bar{h}_1(-x)$ for $x<0$.
On the other hand, integrating $H_T(x, \xi, t)$ over $x$ gives the tensor form factor:
\begin{equation}
\int_{-1}^{1} dx H_T(x, \xi, t) = g_T(t).
\end{equation}
Another feature of GPDs important for our investigations below 
is their interpretation as densities in the transverse plane for $\xi=0$ \cite{Burkardt:2000za}.
To give an example, it has been shown that the impact parameter
dependent quark distribution for the quark GPD $H_q$,
\begin{eqnarray}
q(x,b_\perp)
\equiv\int \frac{d^2\Delta_\perp}{(2\pi)^2} e^{-i b_\perp \cdot \Delta_\perp} H_q(x,\xi=0,t=-\Delta_\perp^2),
\label{Fourier}
\end{eqnarray}
has the interpretation of a probability density for unpolarized quarks of flavor $q$ with longitudinal momentum fraction
$x$ and transverse position $b_\perp=(b_x,b_y)$ relative to the center of momentum in a nucleon.

In order to facilitate the computation of the tensor GPDs in lattice QCD,
we first transform the LHS of Eq.~(\ref{GPDs1}) to Mellin
space by forming the integral $\int_{-1}^{1} dx x^{n-1}\cdots$.
This results in nucleon matrix elements of towers of local tensor
operators
\begin{equation}
{\mathcal O}_T^{\mu\nu \mu_{1}\ldots \mu _{n-1}}(0)= \bar{q}(0)
i\sigma^{\mu \{\nu }\gamma_5 i\Dlr{}^{\mu_{1}}\ldots
i\Dlr{}^{\mu_{n-1}\}}q(0)\, ,
\label{eq:ops}
\end{equation}
which are parametrized in terms of tensor generalized
form factors (GFFs) $A_{Tni}$, $\overline B_{Tni}$, $\widetilde A_{Tni}$ and
$\widetilde B_{Tni}$.
Here, $\Dlr = \frac{1}{2}(\overrightarrow{D} -
\overleftarrow{D})$ and $\{\cdots\}$ indicates symmetrization of
indices and subtraction of traces.\hspace{-1pt}\footnote{The Mellin-moment index $n$ used in this work differs from
  the $n$ in \cite{Hagler:2004yt} by one.}
For $n=1$, we have \cite{Diehl:2001pm,Hagler:2004yt}
\begin{eqnarray}
\left\langle P^{\prime}\Lambda^{\prime}\right| 
 \bar{q}(0) \!\!\!\! && \!\!\!\! i\sigma^{\mu\nu}\gamma_5 q(0)
\left| P \Lambda\right\rangle
=\overline u(P',\Lambda ') \bigg\{ \sigma^{\mu\nu}\gamma_5
   \bigg( A_{T10}(t) - \frac{t}{2m^2}\widetilde A_{T10}(t) \bigg)\nonumber\\
  &+& \frac{\eps^{\mu\nu\alpha\beta} \Delta_{\alpha} \gamma_{\beta}} {2 m} \overline B_{T10}(t)
 +  \frac{\Delta^{[\mu} \sigma^{\nu]\alpha}\gamma_5 \Delta_{\alpha]}} {2m^2 } \widetilde A_{T10}(t)\!\bigg\}
 u(P,\Lambda)\ .
\label{tn0}
\end{eqnarray}
The relation of the lowest moment of the tensor GPDs to the GFFs is simple and given by
\begin{equation}
\begin{array}{lcl}
H^{n=1}_{T}(\xi,t) = A_{T10}(t) = g_T(t), & \qquad &
\widetilde{H}^{n=1}_{T}(\xi,t) = \widetilde{A}_{T10}(t)\nonumber \\
\overline E^{n=1}_{T}(\xi,t) = \overline B_{T10}(t), & \qquad &
\widetilde{E}^{n=1}_{T}(\xi,t) = \widetilde{B}_{T10}(t) = 0\, ,
\end{array}
\label{poly}
\end{equation}
where $H^{n}_{T}(\xi,t)\equiv \int_{-1}^{1} dx x^{n-1}H_{T}(x,\xi,t)$.
The general parametrization in terms of GFFs and their relations
to the moments of the GPDs for $n\ge1$ can be found in \cite{Hagler:2004yt,Chen:2004cg}.

The calculation of moments of GPDs in lattice QCD follows standard methods, which have been described in detail
in the literature \cite{QCDSF-1,MIT,Gockeler:2005cj}. In the following, we therefore give only an outline
of the procedure we use to extract the GFFs. First, nucleon matrix elements in the form of two- and three-point
functions are computed on an Euclidean space-time lattice.
The typical suppression of the matrix elements by exponential factors $\exp(-\tau E)$ in the Euclidean time
$\tau$ and the energy $E$ is cancelled out by constructing an appropriate ratio $R(\tau)$
of three- to two-point functions, which is averaged over the plateau-region
$R(\tau_{\mbox{{\scriptsize plat.}}}) \approx {\mbox{const}}$. The averaged ratio is then renormalized
and equated with the continuum parametrization of the corresponding nucleon matrix element, e.g.\ Eq.~({\ref{tn0}),
for all contributing index ($\mu,\nu$) and momentum ($P,P'$) combinations.
This leads to an overdetermined set of linear equations which is solved to extract the GFFs.
The statistical error on the GFFs is obtained from a jackknife analysis.
Our results have been non-perturbative renormalized \cite{reno}
and transformed to the $\overline{\mbox{MS}}$ scheme at a scale of $4$ GeV$^2$.

The lattice results to be discussed below have been obtained from simulations with 
$\Nf=2$ flavors of dynamical non-perturbatively ${\mathcal O}(a)$ improved Wilson fermions and Wilson glue.
There are 12 datasets available consisting of four different couplings
$\beta=5.20$, $5.25$, $5.29$, $5.40$ with three different $\kappa=\kappa_{\mathrm {sea}}$ values per $\beta$.
The pion masses of our calculation vary from $550$ to $1000$ MeV, and
the lattice spacings and spatial volumes vary between 0.07-0.11~fm and (1.4-2.0~fm)$^3$ respectively.
Our calculation does not include the computationally demanding disconnected contributions. We expect,
however, that they are small for the tensor GFFs \cite{Gockeler:2005cj}.
More details of the simulation can be found in \cite{Gockeler:2005cj,Gockeler:2004mn,Gockeler:2005aw}.


\section{Lattice study of the Soffer bound}
\begin{figure}[t]
\bc
\includegraphics[height=11.7cm,angle=-90]{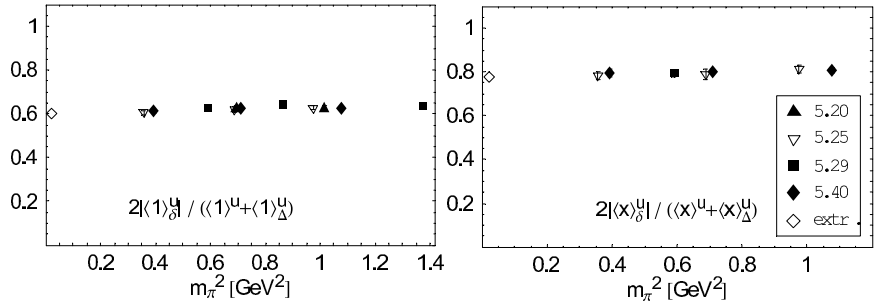}
\caption{Lattice results for the lowest two moments of the Soffer bound for up-quarks.}
\label{soffer1}
\ec
\end{figure}
%
%
In this section, we investigate the Soffer bound
\cite{Soffer:1994ww}
\begin{equation}
\left|\delta q(x) \right|\le \frac{1}{2} \Big( q(x) + \Delta q(x) \Big)\ ,
\label{soffereq}
\end{equation}
which holds exactly only for quark and anti-quark distributions separately.
For discussion of its validity, see e.g. \cite{Altarelli:1998gn} and section 3.12.3 of \cite{Diehl:2003ny}.
For a lattice study of the Soffer bound, we take Mellin moments of
Eq.~(\ref{soffereq}) and consider the ``Soffer-ratio''
\begin{equation}
S^n=\frac{2\left|\langle x^{n-1}\rangle_{\delta}\right|}
     {\langle x^{n-1}\rangle +
                        \langle x^{n-1}\rangle_{\Delta} }\ ,
\qquad n=1,2,\ldots \, ,
\label{sofferratio}
\end{equation}
where $\langle x^{n-1}\rangle=\int^1_{-1} dx x^{n-1}q(x)$.
At this point it is important to note that Mellin moments of distribution functions
give always sums/differences of moments of quark and anti-quark distributions,
e.g.\ $\langle x^{n-1}\rangle=\langle x^{n-1}\rangle_q+(-1)^{n}\langle x^{n-1}\rangle_{\bar q}$. Therefore,
the ratio $S^n$ in Eq.~(\ref{sofferratio}) is not necessarily smaller than one.
Experience shows that contributions from anti-quarks are negligible in our calculation.
In Fig.~(\ref{soffer1}) we show our results for the ratio (\ref{sofferratio}) versus the pion mass
for up-quarks (similar results for down-quarks can be found in \cite{Gockeler:2005cj}).
The fact that the ratio is consistently below one for the lowest two moments of the up and the down
quarks strongly suggests that 
the Soffer bound is
satisfied in our lattice calculation. 
The lattice results show almost no dependence on the pion mass due to cancellations
of the pion mass dependence of the individual distribution functions
in the ratio~(\ref{sofferratio}). 
Linear chiral extrapolation in $m_\pi^2$ leads to the following predictions
for the ratios at the physical pion mass
\begin{equation}
\begin{array}{ll}
 \mbox{up-quarks }&: \,\,\,S^{n=1} = 0.60 \!\pm\! .01 ,\,\, S^{n=2} = 0.78 \!\pm\! .01 \\
 \mbox{down-quarks }&:\,\,\,  S^{n=1} = 0.57 \!\pm\! .02 ,\,\, S^{n=2} = 0.73 \!\pm\! .05 \, .
\label{ratios2}
\end{array}
\end{equation}
\section{Lattice results for the lowest moment of the transverse spin density}
We now turn our attention to a discussion of our lattice results for the density of transversely polarized
quarks in the nucleon. The lowest moment of the quark transverse spin density is given by \cite{Diehl:2005jf}
\begin{eqnarray}
\left\langle P^+,R_\perp=0,S_\perp \right|
\!\!\!& &\!\!\!\frac{1}{2} \overline q(b_\perp)\big[\gamma^+ - s_\perp^j i \sigma^{+j}\gamma_5\big]q(b_\perp)
\left| P^+,R_\perp=0,S_\perp \right\rangle
= \nonumber\\
\!\!\!&&\!\!\!\frac{1}{2}\left\{
  A_{10}(b_\perp)
+ s_\perp^i S_\perp^i \left( A_{T10}(b_\perp)
- \frac{1}{4m^2} \Delta_{b_\perp} \widetilde{A}_{T10}(b_\perp) \right) \right.  \nonumber\\
&+& \frac{ b_\perp^j \eps ^{ji}}{m} \left( S_\perp^i B_{10}'(b_\perp)
+ s_\perp^i \overline{B}_{T10}'(b_\perp) \right) \nonumber\\
&+& \left. s_\perp^i ( 2 b_\perp^i b_\perp^j
- b_\perp^2 \delta^{ij} ) S_\perp^j \frac{1}{m^2} \widetilde{A}_{T10}''(b_\perp)
\right\}.
  \label{density}
\end{eqnarray}
The transversity states
\begin{equation}
\left| P^+,R_\perp=0,S_\perp \right\rangle = 
\frac{1}
{\sqrt{2}}\Big(\left| P^+,R_\perp=0,\Lambda=+\right\rangle + e^{i\chi}\left| P^+,R_\perp=0,\Lambda=-\right\rangle\Big)
\end{equation}
describe a nucleon with longitudinal momentum $P^+=(P^0+P^3)/\sqrt{2}$ which is localized in the transverse
plane at $R_\perp=0$ and has transverse spin $S_\perp = (\cos\chi,\sin\chi)$.
The impact parameter dependent GFFs in Eq.~(\ref{density}) are just the Fourier-transforms
of the momentum space GFFs at $\xi=0$, as in Eq.~(\ref{Fourier}).
\begin{figure}[t]
\bc
\includegraphics[height=10cm,angle=-90]{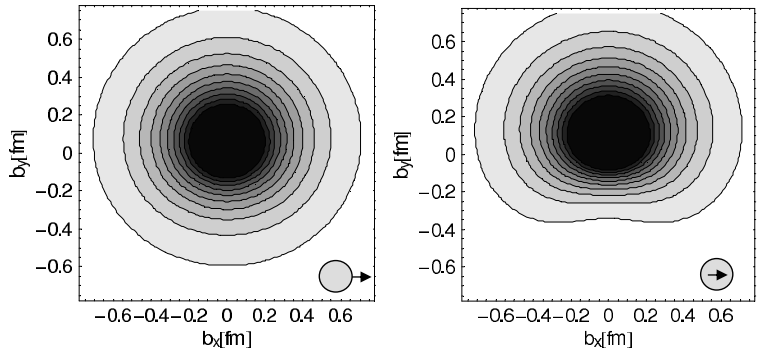}
\caption{Densities of up-quarks in the nucleon.
The nucleon and quark spins are oriented in the transverse plane as indicated, where
the inner arrow represent the quark and the outer arrow the nucleon spin. 
A missing arrow represents the unpolarized case.}
\label{densities1}
\ec
\end{figure}
The derivatives in Eq.~(\ref{density}) are defined by $f'(b_\perp) \equiv \partial_{b_\perp^2}f(b_\perp)$ and
$\Delta_{b_\perp}f(b_\perp) \equiv 4\partial_{b_\perp^2}\big(b_\perp^2\partial_{b_\perp^2}\big)f(b_\perp)$.

Since momenta are discretized on a finite lattice, we obtain the GFFs only for a limited number
of different values of the momentum transfer squared $t$. We have in general 16 $t$-values available
per dataset in a range of $0\le t < 4$ GeV$^2$. To facilitate the Fourier transformation to
impact parameter space, we parametrize the GFFs using a $p$-pole ansatz
\begin{equation}
 F(t)=\frac{F(0)}
  {\left( 1 - {t/m_p^2}
   \right)^p} \ ,
\label{ppole}
\end{equation}
where the parameters $F(0)$, $m_p$ and $p$ for the individual GFFs are fixed by a fit to the lattice results.
The ansatz in Eq.~(\ref{ppole}) is then Fourier transformed in order to get the GFFs as functions
of the impact parameter $b_\perp$.
Details of the $p$-pole parametrization and numerical results for the parameters will be given
in a separate publication \cite{GPDprep}. Here we only note that the values for the power $p$ we are
using for the different GFFs lead to a regular behavior of the transverse spin density
in the limit $b_\perp\to 0$, as discussed in \cite{Diehl:2005jf}.
In Figs.(\ref{densities1},\ref{densities2}) we show preliminary results
for the lowest moment of transverse spin densities of quarks in the nucleon for
up quarks.
We note that the plots do not exactly show probability densities because the lowest moment
corresponds to the difference of quark and anti-quark densities. The densities are however
strictly positive for all $b_\perp$, indicating that the contributions from anti-quarks are small.
On the LHS of Fig.(\ref{densities1}), we show the transversely distorted density of unpolarized quarks in a nucleon
with spin in $x$-direction (coming from the dipole-term $\propto \eps ^{ji} b_\perp^j S_\perp^i$ in Eq.~(\ref{density})), which
has already been discussed in \cite{Burkardt:2002hr}.
A new observation is that the GPD $\overline E_T$ also leads to a strong transverse distortion
orthogonal to the transverse quark spin for an \emph{unpolarized} nucleon (coming from the dipole-term
$\propto \eps ^{ji} b_\perp^j s_\perp^i$ in Eq.~(\ref{density})) 
on the RHS of Fig.(\ref{densities1}). It was argued in \cite{Burkardt:2005hp} that this shift
in $+y$-direction\footnote{or equivalently a shift in $(-x)$-direction for quarks with spin in $y$-direction}
may correspond to a non-zero, negative Boer-Mulders function \cite{Boer:1997nt} $h_1^{\perp}<0$ for up-quarks.
\begin{figure}[t]
\bc
\includegraphics[height=10cm,angle=-90]{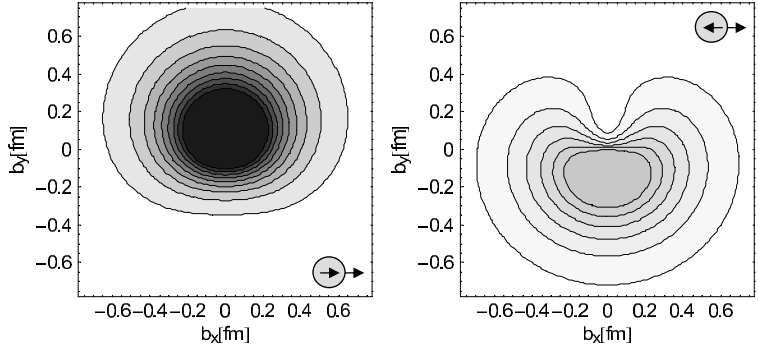}
\caption{Up-quark densities. Symbols are explained in the caption of Fig.(\ref{densities1}).}
\label{densities2}
\ec
\end{figure}
%
The distortions due to transverse quark and nucleon spin add up for the density on the LHS in Fig.(\ref{densities2}),
while it goes in opposite ($-y$)-direction for quarks with spin opposite to the nucleon spin,
as can be seen on the RHS of Fig.(\ref{densities2}).
Interestingly, there is practically no influence visible from the quadrupole-term $\propto s_\perp^i ( 2 b_\perp^i b_\perp^j
- b_\perp^2 \delta^{ij} ) S_\perp^j$ in Eq.~(\ref{density}) for the up-quark densities.
%
%

\section{Conclusions and outlook}

Our lattice results for the transversity distribution
suggests that the Soffer bound is saturated by $\approx 60-80\%$ for the lowest two $x$-moments.
In addition, we have presented preliminary results for the lowest moment of the
transverse spin density of quarks in the nucleon. The distortion of the density of
transversely polarized quarks in an unpolarized nucleon is substantial
and could give rise to a non-vanishing negative Boer-Mulders function for up-quarks
through final state interactions as argued by Burkardt \cite{Burkardt:2005hp}.

We plan to extend our analysis of transverse spin densities in lattice QCD to the lowest two moments
of up- and down-quarks and to investigate improved positivity bounds for GPDs which have
been obtained in \cite{Diehl:2005jf}.

\section*{Acknowledgments}

The numerical calculations have been performed on the Hitachi SR8000
at LRZ (Munich), on the Cray T3E at EPCC (Edinburgh) 
\cite{Allton:2001sk}, and on the APEmille at NIC/DESY
(Zeuthen). This work is supported by the DFG (Forschergruppe
Gitter-Hadronen-Ph\"anomenologie and Emmy-Noether-program), 
by the EU I3HP under contract number RII3-CT-2004-506078 and by the 
Helmholtz Association, contract number VH-NG-004.


\begin{thebibliography}{99}

\bibitem{GPD}
  D.~M\"uller {\it et al.},
  Fortsch.\ Phys.\  {\bf 42} (1994) 101
  [hep-ph/9812448];
  X.~Ji,
  Phys.\ Rev.\ D {\bf 55} (1997) 7114
  [hep-ph/9609381];
  A.~V.~Radyushkin,
  Phys.\ Rev.\ D {\bf 56} (1997) 5524
  [hep-ph/9704207].

\bibitem{Diehl:2003ny}
  M.~Diehl,
  Phys.\ Rept.\  {\bf 388} (2003) 41
  [hep-ph/0307382].

\bibitem{Ji:1996ek}
  X.~Ji,
  Phys.\ Rev.\ Lett.\  {\bf 78} (1997) 610
  [hep-ph/9603249].

\bibitem{Burkardt:2000za}
  M.~Burkardt,
  Phys.\ Rev.\ D {\bf 62} (2000) 071503
  [Erratum-ibid.\ D {\bf 66} (2002) 119903]
  [hep-ph/0005108].
  \bibitem{Diehl:2002he}

  M.~Diehl,
  Eur.\ Phys.\ J.\ C {\bf 25} (2002) 223
  [Erratum-ibid.\ C {\bf 31} (2003) 277]
  [hep-ph/0205208].


  \bibitem{MIT-2}
  Ph.~H\"agler {\it et al.},
  Phys.\ Rev.\ Lett.\ \textbf{93} (2004) 112001
  [hep-lat/0312014].






\bibitem{Diehl:2005jf}
  M.~Diehl, Ph.~H\"agler, Eur. Phys. J. C \textbf{44} (2005) 87
  [hep-ph/0504175].

\bibitem{Diehl:2001pm}  M.~Diehl,
Eur.\ Phys.\ J.\ C \textbf{19} (2001) 485
[hep-ph/0101335].

\bibitem{Burkardt:2005hp}
  M.~Burkardt,
  hep-ph/0505189.

\bibitem{Hagler:2004yt}
Ph.~H\"agler,
Phys.\ Lett.\ B {\bf 594} (2004) 164
[hep-ph/0404138].

\bibitem{Chen:2004cg}
  Z.~Chen and X.~Ji,
  Phys.\ Rev.\ D {\bf 71} (2005) 016003
  [hep-ph/0404276].

\bibitem{QCDSF-1}
  M.~G\"ockeler {\it et al.},
  Phys.\ Rev.\ Lett.\  {\bf 92} (2004) 042002
  [hep-ph/0304249].

  \bibitem{MIT}
  Ph. H\"agler {\it et al.}, Phys. Rev. D \textbf{68} (2003) 034505
  [hep-lat/0304018].

\bibitem{Gockeler:2005cj}
  M.~G\"ockeler {\it et al.}, Phys. Lett. B {\bf 627} (2005) 113
  [hep-lat/0507001].

\bibitem{reno}
  G.~Martinelli {\it et al.},
  Nucl.\ Phys.\ B {\bf 445} (1995) 81
  [hep-lat/9411010];
  M.~G\"ockeler {\it et al.},
  Nucl.\ Phys.\ B {\bf 544} (1999) 699
  [hep-lat/9807044].

\bibitem{Gockeler:2004mn}
  M.~G\"ockeler {\it et al.},
  Few Body Syst.\  {\bf 36} (2005) 111
  [hep-lat/0410023].

\bibitem{Gockeler:2005aw}
  M.~G\"ockeler {\it et al.},
  Nucl.\ Phys.\ A {\bf 755} (2005) 537
  [hep-lat/0501029].



\bibitem{Soffer:1994ww}
  J.~Soffer,
  Phys.\ Rev.\ Lett.\  {\bf 74} (1995) 1292
  [hep-ph/9409254];
  D.~W.~Sivers,
   Phys.\ Rev.\ D {\bf 51} (1995) 4880.
   
\bibitem{Altarelli:1998gn}
  G.~Altarelli {\it et al.},
  Nucl.\ Phys.\ B {\bf 534} (1998) 277
  [hep-ph/9806345].
  

    \bibitem{GPDprep}
  M.~G\"ockeler {\it et al.}, {\it in preparation}.
  

\bibitem{Burkardt:2002hr}
  M.~Burkardt,
  Int.\ J.\ Mod.\ Phys.\ A {\bf 18} (2003) 173
  [hep-ph/0207047].

\bibitem{Boer:1997nt}
  D.~Boer, P.~J.~Mulders,
  Phys.\ Rev.\ D {\bf 57} (1998) 5780
  [hep-ph/9711485].
  
\bibitem{Allton:2001sk}
  C.~R.~Allton {\it et al.},
  Phys.\ Rev.\ D {\bf 65} (2002) 054502
  [hep-lat/0107021].

\end{thebibliography}
\end{document}